\documentstyle[12pt]{article}
\begin{document}
\parskip=6pt
\baselineskip=20pt
{\raggedright AS-ITP-94-62}
\smallskip

{\raggedleft November, 1994}
\bigskip
\smallskip

\bigskip
\centerline{\Large\bf  Cyclic Quantum Dilogarithm and Shift Operator \footnote
{This research was partially supported by the China center of advanced science
and technology.}}
\vspace{4ex}
\smallskip
\centerline{\large\bf Zhan-Ning Hu \footnote{\bf email address:
huzn@itp.ac.cn}}
\smallskip
\centerline{CCAST(WORLD LABORATORY) P.O.BOX 8730, BEIJING, 100080}
\centerline{and}
\centerline{INSTITUTE OF THEORETICAL PHYSICS, ACADEMIA SINICA,}
\centerline{ P. O. BOX 2735 BEIJING 100080, CHINA \footnote{\bf mail address}}
\vspace{2ex}
\centerline{\large\bf Bo-Yu Hou}
\centerline{INSTITUTE OF MODERN PHYSICS, NORTHWEST UNIVERSITY, XIAN, CHINA}
\smallskip

\vspace{4ex}
\bigskip
\begin{center}
\begin{minipage}{5in}
\centerline{\large\bf 	Abstract}
\vspace{3ex}
{}From the cyclic quantum dilogarithm the shift operator is constructed with
$q$ is a root of unit and the representation is given for the current algebra
introduced by Faddeev $et ~al$. It is shown that the theta-function is
 factorizable also in this case by using the star-square equation of the
Baxter-Bazhanov model.

\vspace{3ex}



\end{minipage}
\end{center}
\newpage

\section{\bf Introduction}
Recently people have pay more attention to the investigation of completely
integrable discrete variants of conformal field theory such as lattice
Wess-Zumino-Witten model \cite{fer}, quantum Volterra model \cite{vol1}, the
lattice Liouville model and the lattice sine-Gordon model \cite{fadtak}.
 Bruschi $et~ al$ studied the properties and behaviors of integrable symplectic
maps from the lattice evolution equation\cite{bru}. In order to providing the
 new way to construct the conservation laws and to discuss the nature of
quantum integrabilit by using the discrete space-time method
\cite{desveg,cappap} Faddeev $et~ al$ investigated the lattice Virasoro
algebra and the corresponding hierarchy of conservation laws from the lattice
 current algebra where the shift operator with $|q|<1$ is discussed in
detail \cite{fv}.

We know that the dilogarithm appeared firstly in the discussion of the
 $XXZ$-model at small magnetic field \cite{BR}, then appeared in the
 2-dimensional quantum filed theories and solvable lattice models \cite{many}.
 The quantum dilogarithm identity introduced in Ref. \cite{fk} is equivalent
 to the restricted star-triangle relation of the three-dimensional
 Baxter-Bazhanov model \cite{bb12,old1} and more recently Kashaev built a
 connection between cyclic $6j$-symbol and the quantum dilogarithm.

In this letter, we will construct the shift operator from the cyclic quantum
dilogarithm and give a representation of the lattice current algebra which
 combined with free discrete time dynamics. It is proved that the
``theta-function'' is also factorizable in the case of $|q|=1$ by using the
star-square equation \cite{k2,h2} of the three-dimensional Baxter-Bazhanov
 model. In section 2 we give the cyclic quantum dilogarithm and the
representation of the lattice current algebra. The shift operator is
constructed and the "theta-function" is proved to be factorizable in section 3.
Finally some remarks are given.

\section{\bf Cyclic Quantum Dilogarithm and the Lattice Current Algebra}
By following the Refs.\cite{bb12,k2}, define the function
\begin{equation}
w(a,b,c|l)=\prod^l_{j=1}b/(c-a\omega^j),~~a^L+b^L=c^L,~~l\geq 0,
\end{equation}
with $w(a,b,c|0)=1$. For any operator $A$ whose $L$-th power is the identity
 operator, the spectrum of the operator $A$ is given by $L$ distinct numbers
\begin{equation} \label{spe}
\omega^l,~~l=0,1,\cdots,L-1.
\end{equation}
We define the cyclic quantum dilogarithms $\Psi(A)$ and $\Phi(A)$ (See Ref.
 \cite{fk}) are the commuted operators which  depend on the operator $A$ and
commute with $A$ which has the spectrum (\ref{spe}). The spectrums of $\Psi(A)$
and $\Phi(A)$ have the following forms:
$$
\Psi(\omega^l)=\Psi(1)w(a,b,c|l), ~~~~~~~
$$
\begin{equation}
\Phi(\omega^l)=\Phi(1)w(c,\omega^{1/2}b,\omega a|l),
\end{equation}
where $\Psi(1)$ and $\Phi(1)$ are the non-zero complex factors. And the
 ``functional'' relations of the cyclic quantum dilogarithms $\Psi(A)$ and
$\Phi(A)$ can be written as
$$
\Psi(\omega^{-1}A)\Psi(A)^{-1}=(c-aA)/b, ~~~~~~~~
$$
\begin{equation} \label{fun}
{}~~~~~ \Phi(\omega^{-1}A)\Phi(A)^{-1}=(\omega^{1/2}a-\omega^{-1/2}cA)/b,
\end{equation}
which determine the operators $\Psi(A)$ and $\Phi(A)$ up to the complex
factors. Furthermore, from the above relations, we have
\begin{equation}
\Psi(A)=\rho_1\sum^{L-1}_{l=0}A^l\prod^l_{j=1}\frac{a}{c-b\omega^{-j}}, ~~~~~~
\end{equation}
\begin{equation}
\Phi(A)=\rho_2\sum^{L-1}_{l=0}A^l\prod^l_{j=1}\frac{c}{a\omega-b\omega^{1/2-j}},
\end{equation}
where $\rho_1$ and $\rho_2$ are also the non-zero complex factors. In this way,
 $\Psi(1)$ and $\Phi(1)$ can be expressed as
$$
\Psi(1)=\rho_1\frac{c-a\omega}{b}\sum^{L-1}_{l=0}\prod^l_{j=1}\frac{a\omega}
{c-b\omega^{-j}}, ~~~~~~~~~~~~~~~
$$
\begin{equation}
\Phi(1)=\rho_2\frac{\omega^{1/2}(a-c)}{b}\sum^{L-1}_{l=0}\prod^l_{j=1}
\frac{c\omega}{a\omega-b\omega^{1/2-j}}.
\end{equation}
By using the cyclic quantum dilogarithms (\ref{fun}) we will construct the
shift operator with $|q|=1$, in the following section, from the lattice
 current algebra
$$
w_{n-1}w_n=q^2w_nw_{n-1},~~~n=2,3,\cdots,2N,
$$
\begin{equation} \label{1}
w_{2N}w_1=q^2w_1w_{2N},
\end{equation}
$$
w_mw_n=w_nw_m, 1<|m-n|<2N-1,
$$
which reduces to the periodic free field in the continuous limit \cite{fv}.
Set
$$
x_i=\underbrace{I\otimes I\otimes \cdots \otimes I}_{i-1}\otimes
 x\otimes\underbrace{I\otimes I\otimes \cdots \otimes I}_{N-i-1},
$$
\begin{equation}
y_i=\underbrace{I\otimes I\otimes \cdots \otimes I}_{i-1}\otimes
 y\otimes\underbrace{I\otimes I\otimes \cdots \otimes I}_{N-i-1},
\end{equation}
with $i=1,2,\cdots,N-1$, where the $L$-by-$L$ matrices $x, y$ are given by
\begin{equation}
\begin{array}{cc}
x=\left[
\begin{array}{ccccccc}
0&1&0&0&\cdots &0&0\\
0&0&1&0&\cdots &0&0\\
\vdots&\vdots&\vdots&\vdots&\cdots&\vdots&\vdots\\
0&0&0&0&\cdots &0&1\\
1&0&0&0&\cdots &0&0
\end{array}
\right], ~&
y=\left[
\begin{array}{ccccc}
1&0&0&\cdots&0\\
0&\omega&0&\cdots&0\\
0&0&\omega^2&\cdots&0\\
\vdots&\vdots&\vdots&\cdots&\vdots\\
0&0&0&\cdots&\omega^{L-1}
\end{array}
\right],~~
\end{array}
\end{equation}
with $\omega=exp(2\pi i/L)$ and $I$ is the $L$-by-$L$ unit matrix. When we fix
that $w^L_n=1 ~(n=1,2,\cdots,2N)$ the representation of the lattice current
algebra (\ref{1}) with $q^2=\omega$ is given from the following relations:
\begin{equation}
\begin{array}{l}
w_{2i+1}=x^{-1}_ix_{i+1},~~i=0,1,2,\cdots,N-1, \\
w_{2j}=y_j,~~j=1,2,\cdots,N-1, \\
w_{2N}=\prod^{N-1}_{l=1}y^{-1}_l,
\end{array}
\end{equation}
where $x_0=x_N=1$ and $x_i, y_i (i=1,2,\cdots,N-1)$ are the $N-1$ independent
Weyl pairs satisfied the relations $x_iy_i=q^2y_ix_i$.

\section{\bf Shift Operator and Factorized ``Theta Function'' }
In the discrete space-time picture the equation of motion of the free field
 $p(x)$, $p_t(x,t)+p_x(x,t)=0$, has the form $w_n(t+1)=w_{n-1}(t)$ with
 $w_n(0)=w_n$. Then it exists that the shift operator which satisfy
\begin{equation} \label{2.1}
w_nU=Uw_{n-1}.
\end{equation}
The main aim in the rest of this letter is to discuss it by using the cyclic
 quantum dilogarithms. Let
\begin{equation} \label{hn}
h_n=\Psi(w_n)\Phi(w^{-1}_n)
\end{equation}
with $w^L_n=1$ for $n=1,2,\cdots,2N$. From relation (\ref{fun}) we have
\begin{equation}
w_nh_{n-1}=-\omega^{1/2}h_{n-1}w_nw_{n-1},~~~
h_{n}w_{n-1}=-\omega^{1/2}w_nw_{n-1}h_n.
\end{equation}
It can be obtained easily that
\begin{equation} {\label w}
w_nh_{n-1}h_n=h_{n-1}h_nw_{n-1},~~n=1,2,3,\cdots,2N,
\end{equation}
with $w_0=w_{2N}$. By considering that the two central elements $C_1=\prod
w_{odd}$ and $C_2=\prod w_{even}$ are equal \footnote{We can always get the
lattice current algebra with the two equal central elements $C_1$ and $C_2$ by
choosing $w_n$ properly.}, from the above relation, we can express the shift
operator $U$ as
\begin{equation} {\label u}
U=h_1h_2\cdots h_{2N-1},
\end{equation}
which satisfies the relation (\ref{2.1}). And operators $h_n$ give a
representation of the braid group of the $A^{(1)}_{2N-1}$ type. It is
interesting that the above relations hold also when we substitute operators
$h_n$ by $\bar{h}_n=\Phi(w_n)\Psi(w^{-1}_n)$. In the other hand, the shift
operator with $|q|\not= 1$ can be denoted by the theta functions
\cite{fk,hunew}
. Then we can ask the questions that what is the difference between the
operators $h_n$ and $\bar{h}_n$ and what happened about the theta functions
when it denotes the shift operator with $|q|=1$. The answers of them are given
as the follows. We know that the star-square relation of the three dimensional
 Baxter-Bazhanov model can be written as \cite{k2,h2}
$$
\Bigg\{\sum_{\sigma\in Z_N}\frac{w(x_1,y_1,z_1|a+\sigma)
w(x_2,y_2,z_2|b+\sigma)}{w(x_3,y_3,z_3|c+\sigma)w(x_4,y_4,z_4|d+\sigma)}
\Bigg\}_0 ~~~~~~~~~~~~~~~~~~~~~~~~~~~~~~~~~~~~~~~~~~~
$$
$$
=\frac{(x_2y_1/x_1z_2)^a(x_1y_2/x_2z_1)^b(z_3/y_3)^c(z_4/y_4)^d}{\gamma
(a-b)\omega^{(a+b)/2}} ~~~~~~~~~~~~~~~~~~~~~~~~~~~~~~~
$$
\begin{equation} \label{e24}
{}~\times \frac{w(\omega x_3x_4z_1z_2/x_1x_2z_3z_4|c+d-a-b)}
{w\bigg(\frac{\displaystyle x_4z_1}{\displaystyle x_1z_4}|d-a\bigg)
w\bigg(\frac{\displaystyle x_3z_2}{\displaystyle x_2z_3}|c-b\bigg)w\bigg(
\frac{\displaystyle x_3z_1}{\displaystyle x_1z_3}|c-a\bigg)w\bigg(
\frac{\displaystyle x_4z_2}{\displaystyle x_2z_4}|d-b\bigg)},
\end{equation}
with the constraint condition $y_1y_2z_3z_4/(z_1z_2y_3y_4)=\omega$ where the
subscript "0" after the curly brackets indicates that the l. h. s. of the
above equation is normalized to unity at zero exterior spins and the following
notations are used:
\begin{equation}
w(x,y,z|l)=(y/z)^lw(x/z|l), ~ \gamma (a-b)=\omega^{(a-b)(L+a-b)/2}.
\end{equation}
{}From Eqs. (5), (6) and (7), the operators $h_n$ can be denoted as
\begin{equation}
h_n=\rho_1\rho_2\sum^{L-1}_{k=0}a_kw^k_n
\end{equation}
where
\begin{equation}
a_k=\sum^{L-1}_{l=0}\frac{1}{w(\omega^{-1}b,a,c|l)w(\omega^{-1/2}b,c,\omega
a|k+l)}~.
\end{equation}
By using the above star-square relation it can be proved that
\begin{equation} \label{lll}
a_k=\gamma(k)a_0
\end{equation}
where $\gamma(k)$ is given in Eq. (18). Then we have
\begin{equation}
 h_n=a_0\rho_1\rho_2\theta(w_n),~~ \theta(w_n)=\sum^{L-1}_{k=0}(-)^k
\omega^{k^2/2}w_n^{~k}.
\end{equation}
Furthermore we can proved that $h_n=\bar{h}_n$ by using Eqs. (17) and
 (\ref{lll}). Therefore the shift operator is constructed from the cyclic
 quantum dilogarithms by which the ``theta function'' is factorized  when
 $|q|=1$.

\section{\bf Conclusions and Remarks}
periodic free field in the discrete time-space picture is constructed from the
 lattice current algebra for which the cyclic representation is given from
$N-1$ independent Weyl pairs. And we show that the ``theta function'' is also
factorizable when $|q|=1$ by using the star-square relation of the three
dimensional Baxter-Bazhanov model.

In fact, the ``shift'' property can be connected to many domains of physics
such as the quantum field theory, the solvable models of statistical mechanics
 and the dynamical systems. Recently Faddeev and Volkov discussed the shift of
the saw $S$ corresponding to the solution  of Hirota Equation along the
 discrete time axis as an example of an integrable symplectic map \cite{favo}.
 The shift operator also appeared in the differential forms when we set
 $\partial_rf=f-R_rf,~ R_rf(k)=f(k+1),$ then $f(k+1)X=Xf(k)$ on the lattice
 where $X$ is 1-form and it will be useful in the lattice gauge theory
\cite{review}.  Another line of the developments is in the domain of the
 supermanifolds and it is not very clear about the structure of it when the
 relations similar to the lattice current algebra is intrduced \footnote{One
 of the authors (Hu) would like to thank Prof. D. A. Leites, University of
 Stockholm, for the report about supermanifolds in Nov, 1994.}. Due to the
cyclic quantum dilogarithms have the deep connections with the solved lattice
 models we can discuss the shift properties of the two and three dimensional
 lattice model further.

\section*{\bf Acknowledgment}
One of the authors (Hu) would like to thank H. Y. Guo, K. Wu for the helpful
 discussions and B. Yua. Hou, Z. Q. Ma, K. J. Shi, X. C. Song , S. Y. Zhou and
 P. Wang for the interests in this job at the workshop of CCAST(World
 Laboratory) in Beijing.

\end{document}